\documentclass[prb,preprint]{revtex4-1} 

\usepackage{amsmath}  
\usepackage{amsfonts} 
\usepackage{graphicx} 
\usepackage{amssymb}
\begin{document}


\title{Alice and the Foucault Pendulum: the land of action-angle variables}

\author{N. Boulanger}
\email{nicolas.boulanger@umons.ac.be} 
\affiliation{Service de Physique de l'Univers, Champs et Gravitation, Universit\'e de Mons - UMONS, Research Institute for Complex Systems, Place du Parc 20, 7000 Mons, Belgium}

\author{F. Buisseret}
\email{buisseretf@helha.be; fabien.buisseret@umons.ac.be}
\affiliation{CeREF, HELHa, Chauss\'ee de Binche 159, 7000 Mons, Belgium\\ and \\ Service de Physique Nucl\'eaire et Subnucl\'eaire, Universit\'e de Mons - UMONS, Research Institute for Complex Systems, Place du Parc 20, 7000 Mons, Belgium}


\date{\today}

\begin{abstract}
Since the pioneering works of Newton $(1643-1727)$, Mechanics has been constantly 
reinventing itself: reformulated in particular by Lagrange $(1736-1813)$ then Hamilton 
$(1805-1865)$, it now offers powerful conceptual and mathematical tools for the exploration 
of the most complex dynamical systems, essentially via the action-angle variables formulation 
and more generally through the theory of canonical transformations. We give the 
reader an overview of these different formulations through the well-known example of Foucault's 
pendulum, a device created by Foucault $(1819-1868)$ and first installed in the Panth\'eon 
(Paris, France) in 1851 to display the Earth's rotation. The apparent simplicity of the Foucault 
pendulum is indeed an open door to the most contemporary ramifications of Classical 
Mechanics. 
We stress that the action-angle variable formalism is \emph{not} necessary to 
understand Foucault's pendulum. The latter is simply taken as well-known
simple dynamical system used to exemplify modern concepts that \emph{are} 
crucial in order to understand more complicated dynamical  systems. 
The Foucault pendulum installed in the collegiate church of Sainte-Waudru 
(Mons, Belgium) will allow us to numerically estimate the different quantities introduced. 
A free adaptation of excerpts from``Alice's Adventures in Wonderland" will offer 
the reader some poetic breaths. 
\end{abstract}

\maketitle 

 \textsl{Alice was beginning to get very tired of sitting by her sister on the bank, and of having nothing 
 to do. Once or twice she had peeped into the book her sister was reading [: ``Foucault's Pendulum 
 and the various formulations of Mechanics"], but it had no pictures or conversations in it, ``and what 
 is the use of a book," thought Alice, ``without pictures or conversations?" (\dots) suddenly a White 
 Rabbit with pink eyes ran close by her. (\dots) she ran across the field after it and was just in time to 
 see it pop down a large rabbit-hole, under the hedge. In another moment, down went Alice after it. 
 The rabbit-hole went straight on like a tunnel for some way and then dipped suddenly down, so 
 suddenly that Alice had not a moment to think about stopping herself before she found herself 
 falling down what seemed to be a very deep well} \cite{alice}.

 \begin{center}
 	$\backsim$
 \end{center}

\section{Introduction: Sainte-Waudru's Pendulum} 

\begin{figure}[ht]
	\centering
	\includegraphics[width=0.8\textwidth]{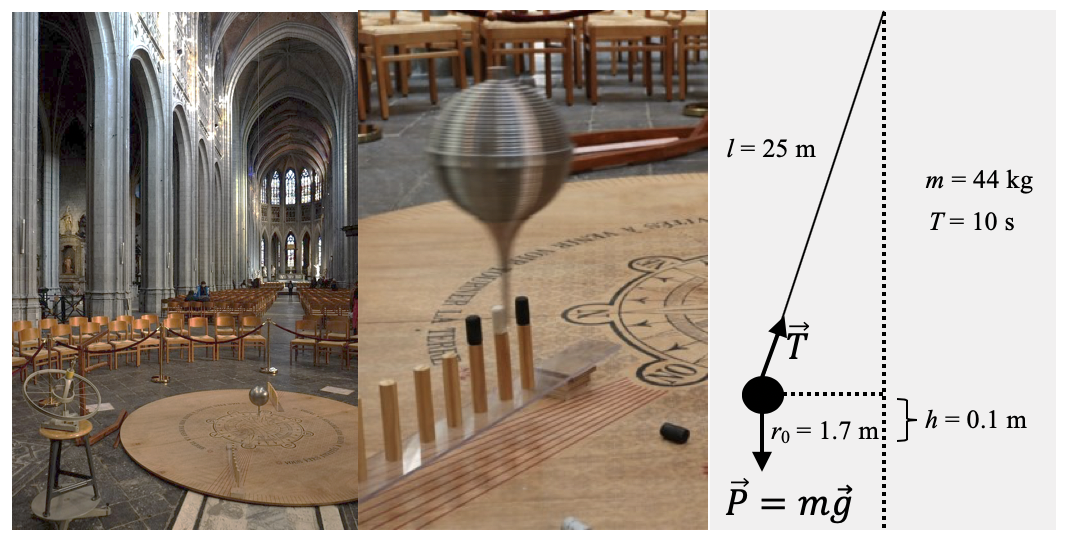}
	\caption{Left: Foucault's pendulum installed in the nave of Sainte-Waudru's collegiate church. Middle: Zoom on the mass of the pendulum. The rotation of the plane of swing of the pendulum
	 is exemplified by the successive fall of the neighbouring corks. Right: Schematic view of the latter pendulum at release. The constant $l$ is the pendulum length. The amplitude is $r_0\,$, causing a vertical displacement $h$ with respect to equilibrium position. After release, the oscillation period is 
	 $T\,$. The vectors $\vec P$ and $\vec T$ are the weight of the mass and the tension of the cable, respectively.}
	\label{fig1}
\end{figure}

The simple pendulum is an easy device to design: it consists of a mass $m$ attached 
at one end of a rigid cable whose mass is negligible compared 
to $m\,$. The other end of the cable is attached at the vertical, thereby suspending the bob of 
the Pendulum. 
It can therefore be considered that the dynamics of the mass $m$ is governed by  Newton's 
equations $\vec T+ \vec P =m\vec a$, $\vec P$ and $\vec T$ being the weight of the bob and 
the tension of the cable, respectively, and $\vec a=\ddot{\vec x}$ the acceleration of the bob 
where $\dot f$ denotes the time derivative of the dynamical variable $f(t)\,$. 
A schematic representation of a simple pendulum is given in Fig. \ref{fig1} and particularised to 
the one installed in the collegiate church of Sainte-Waudru (Mons, Belgium). This Foucault 
pendulum (FP) was installed by the University of Mons (UMONS) in 2005 and has regularly 
been displayed since then.\cite{SW}

As shown in any textbook on Classical Mechanics, see for example 
\cite{Kibble,LLV1,Arnold,Spindel}, the resolution of Newton's equations 
reveals that the simple pendulum, once slightly set out of its equilibrium position $(r_0\ll l)$, 
performs a periodic swing with period $T=\frac{2\pi}{\omega}\,$, where 
$\omega=\sqrt{\frac{g}{l}}\,$. 
For small oscillations, the period is thus independent both of the mass of the bob and of 
the amplitude $r_0\,$ of the swing. For finite oscillations, the period does depend on the amplitude. 
For some original references, see e.g. \cite{Chessin1895,MacMillan1915,Noble1952}.
Detailed explanations including a discussion of the special case of the motion at the equator 
can be found  in \cite{Somerville1972}.  
For other relevant references, see e.g. \cite{Hart1987,Phillips2004}. 
A precise historical account of Foucault's experiment together with 
references to the various attempts at a theoretical understanding can be found 
in the recent review \cite{Sommeria2017}.

\begin{figure}[ht]
	\centering
	\includegraphics[width=0.45\textwidth]{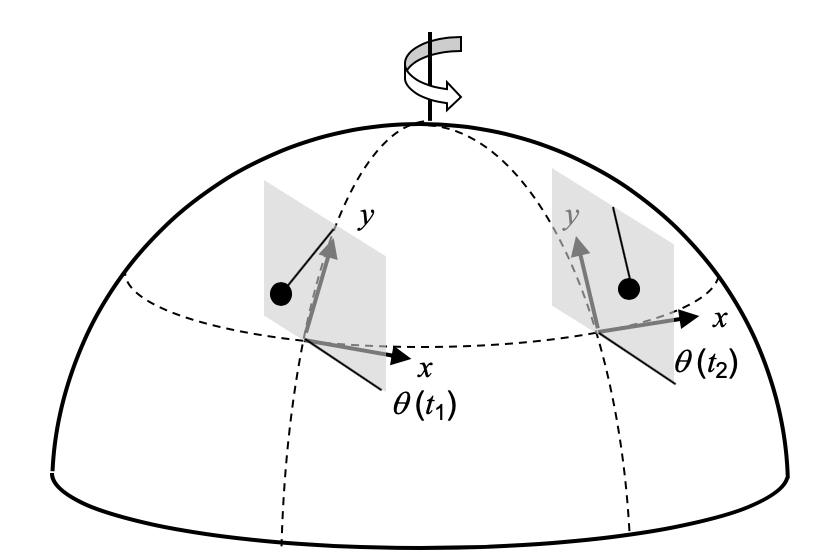}
	\caption{Left: Foucault's pendulum in the fixed reference frame $S_i\,$. 
	Changes in the angle ($\theta$) between the oscillation plane and the $x$ axis of local 
	frame $(x,y)$ are shown at times $t_1$  and $t_2=t_1+\delta t\,$. Right: Sectional view 
	of the Earth and Foucault's pendulum. Pendulum's latitude, $\varphi\,$, is shown as well as 
	the Earth angular velocity $(\vec \Omega_E$) and the oscillation plane's angular velocity 
	($\vec\Omega$).
	}
	\label{fig2}
\end{figure}

 The amplitude sets the total energy of the pendulum, $E$, which can be expressed as the 
 potential energy of the blob at release: $E=mgh\,$. For Saint Waudru's FP one has 
 $\omega=0.626$  rad/s, $T=10$ s and $E=25$ J. Numerical quantities will be given with 
 three significant digits.
 Strictly speaking, frictional forces must be added to the model but they will be neglected 
 here. 
 Friction dissipates the energy of the system: it does not influence the period and only causes 
 a progressive decrease of the amplitude. However, only the periodic behaviour of pendular 
 dynamics is relevant for our purpose.  
 \begin{center}
	$\backsim$
\end{center} 
\textit{``I must be getting somewhere near the centre of the Earth. Let me see: that would be four thousand miles down, I think -" (\dots) ``- yes, that's about the right distance - but then I wonder what Latitude or Longitude I've got to ?" (Alice had not the slightest idea what Latitude was, or Longitude either, but she thought they were nice grand words to say).
}
 \begin{center}
	$\backsim$
\end{center}
In spite of the damping due to friction forces, a FP can oscillate for more than enough time to 
prove \dots that the Earth is rotating! Earth's surface rotates with an angular velocity of 
$\Omega_E=1$ lap/day  with respect to an imaginary fixed sphere (relative to distant stars) $S_i$ of 
same radius and centre as that of the Earth. With respect to $S_i$ however, the normal to the 
instantaneous plane of oscillation of the FP defines an inertial direction: as Newton's Mechanics shows, 
it is a consequence of the fact that the force undergone by the bob of mass $m$ is always directed 
towards the centre of the Earth: there is no sideways force on the bob as viewed from $S_{i}$;   
see for example the extensive discussion in \cite{Somerville1972}. 
This causes an apparent rotation of the oscillation plane viewed in the local reference frame, 
see Fig. \ref{fig2}. 
Only the vertical component of $\vec\Omega_E$ (whose norm is $\Omega=\Omega_E\, \sin\varphi$) 
in the local frame causes this rotation. In our case, FP's latitude  is $\varphi= 50.5^{{\rm o}}$ and 
$\Omega=0.772$ lap/day=5.62 $10^{-5 }$ rad/s. 

\begin{figure}[ht]
	\centering
	\includegraphics[width=0.4\textwidth]{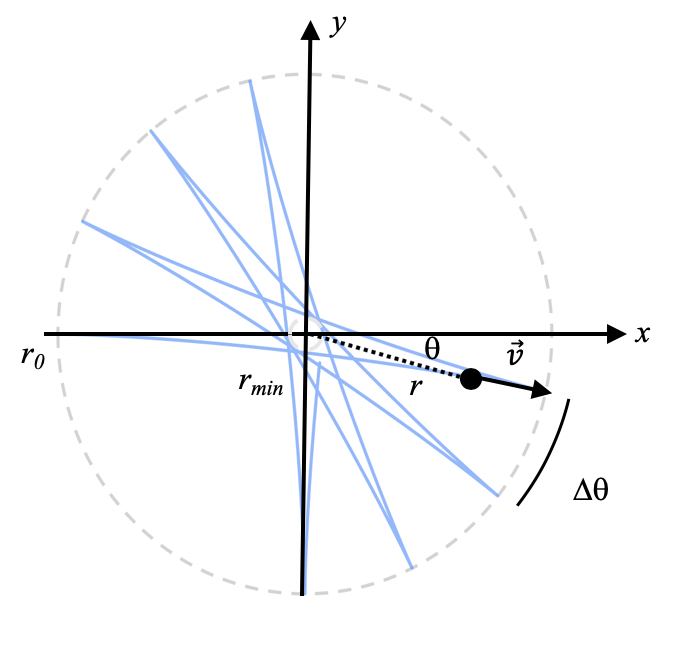}
	\caption{Typical trajectory of the FP in the horizontal plane (the pendulum is thus seen from above). 
The position of the bob of mass $m$ is expressed in polar coordinates $(r,\theta)\,$. 
Pendulum's vertical motion is negligible. 
The shift $\Delta\theta$ and the instantaneous velocity $\vec v$ of the mass are displayed. }
	\label{fig3}
\end{figure}

Newton's equation leads, after calculations, to the radial trajectory 
\begin{equation}\label{eq1}
r(t)=\sqrt{\cos^2(\omega t)+\frac{\Omega^2}{\omega^2}\sin^2(\omega t)}\; .
\end{equation}
The angle $\theta(t)$ may be obtained from $r(t)$ to get the trajectory $(r(t),\theta(t))$ 
in polar coordinates, see \textit{e.g}. \cite{Noble1952}. 
The trajectory in the horizontal plane is a hypocycloid (Fig. \ref{fig3}). 
It can be computed from (\ref{eq1}) that FP never turns back to 
its equilibrium position ($r=0$): there is a minimal radius $r_{min}=\frac{\Omega}{\omega}\, r_0\,$, 
whose origin is Earth's rotation. However, observing this minimum radius is not easy: in Sainte-Waudru it is only 0.153 mm! Another way is more promising: it appears that $\theta(t)$ is shifted by
\begin{equation}\label{eq2}
\Delta\theta=\Omega \,T
\end{equation}
during one period \cite{Noble1952}, that is 0.0322$^{{\rm o}}$ in our example. 
This remains very small, but just wait 10 minutes in Saint-Waudru's nave and the angle of deviation will 
be 1.93$^{{\rm o}}$, which corresponds to a perfectly observable displacement of 5.73 cm on a circle of 
1.7 m radius. This displacement is made visible by the successive falls of the regularly 
spaced corks (see Fig. \ref{fig1}). 

 \begin{center}
	$\backsim$
\end{center}
\textsl{Alice felt that she was dozing off, when suddenly, thump! thump! down she came upon a heap of sticks and dry leaves, and the fall was over. Alice was not a bit hurt, and she jumped up in a moment. She looked up, but it was all dark overhead; before her was another long passage and the White Rabbit was still in sight, hurrying down it. [She followed it to the Hatter's dwelling; after brief greetings, all three of them argued about the curious hypocycloid, and Alice was astonished at the fixed oscillation plane.] ``Have you guessed the riddle yet?" the Hatter said, turning to Alice again.}

\section{Parallel transport along a sphere}

The FP illustrates an important geometric concept called \textit{parallel transport}. 
In this very context, see \cite{Oprea}. 
The velocity of the bob of mass $m$ in the plane tangent to the Earth is a vector. Like any velocity it 
represents an instantaneous displacement along a curve which traces the trajectory. 
The apparent rotation of the oscillation plane can be thought of equivalently as a change 
in the direction of the velocity vector of the bob. This new picture hides an unsuspected difficulty. 
If it is true that in the Euclidean space of Newtonian Mechanics one can compare two vectors 
at different points by dragging them parallel to themselves so as to bring them back to the same point, 
it is however not possible, in a more general space, to compare vectors at different points. 
The space to which the plane $(r,\theta)$ is tangent is nothing more than the Earth, i.e., 
a sphere. As can be seen from Fig. \ref{fig2}, the local plane tangent to the Earth at instant 
$t_1$ is not the one at $t_2$: both planes are tangent to $S_i$, albeit not at the same points.

\begin{figure}[ht]
	\centering
	\includegraphics[width=0.8\textwidth]{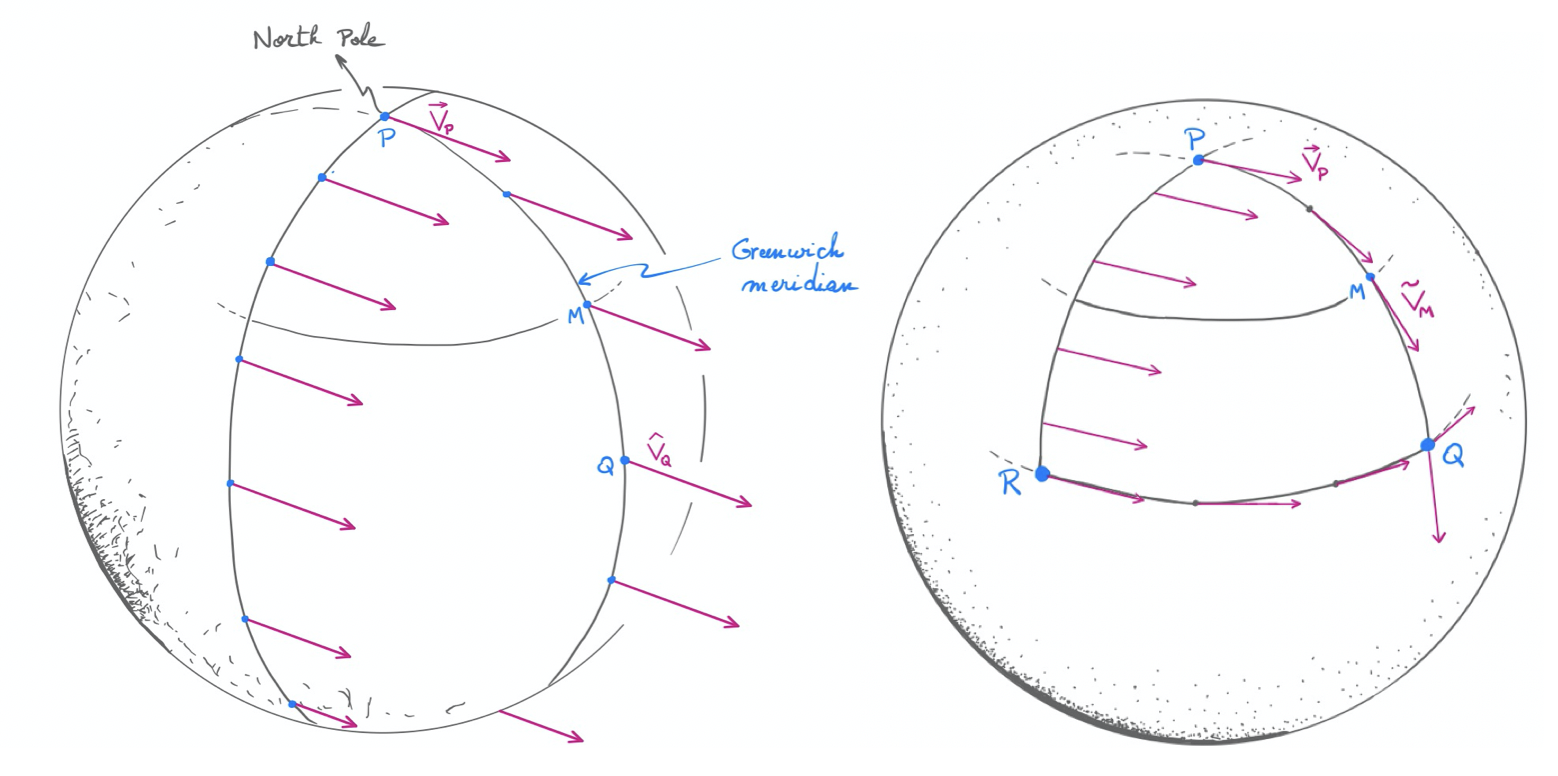}
	\caption{Left: Parallel transport of vector $\vec V_P$ along two meridians computed in the three-dimensional Euclidean space. Right: Parallel transport of vector $\vec V_P$ computed with Levi-Civita's method for the curves PRQ and PQ. }
	\label{fig4}
\end{figure}

How to compare vectors at different points in a sphere? Levi-Civita came up with a method valid for 
general surfaces, that we summarize along the lines of \cite{penrose}. To transport a vector 
$\vec V_P$ from a point P to another point Q of $S_i$, Levi-Civita proposed to embed the sphere into 
Euclidean space and to move the vector $\vec V_P$  from its initial point P to the neighbouring point 
Q using the Euclidean notion of parallel displacement, the very notion of ``sliding" used by our 
schoolchildren in Newtonian Mechanics. The resulting vector, $\hat V_Q\,$, is generally not tangent 
to the sphere $S_i\,$. It has then to be projected onto the surface of the sphere, thereby obtaining a 
vector in Q , denoted $\tilde V_Q\,$. Starting from this new vector and proceeding in a similar way, 
step by step for all the points along a curve passing through P and Q, a whole family of vectors 
tangent to the sphere $S_i$ can be constructed along the latter curve. This construction is 
illustrated in Fig. \ref{fig4}.

Without any specification of a particular curve, the parallel transport operation is ambiguous. 
It is interesting to see what happens for two different curves that connect points P and Q on the Earth's 
surface. Taking the curves PQ and PRQ in Fig. \ref{fig4}, depending on whether one follows the first or 
the second curve, the vector $\tilde V_Q$ obtained by parallel transport will point to the south or will be 
tangent to the equator. Two different orthogonal vectors are thus obtained. Parallel transport along all the 
curves that start from a point and return to the same point gives vectors at the end point that actually 
differ from the initial vector by a rotation by an arbitrary angle. In the case of the Foucault pendulum, the 
relevant curve is the parallel passing through the vertical of its equilibrium position (the parallel of latitude 
50.5$^{{\rm o}}$ in our example). The parallel transport of the body's velocity vector along that parallel 
generates an apparent rotation of its direction for a local observer. This rotation is measurable through 
the displacement of the oscillation plane of the pendulum. A more detailed calculation based on spherical 
trigonometry leads to Eq. (\ref{eq2}).\cite{berg}

\begin{center}
	$\backsim$
\end{center}
\textsl{[Alice was still concerned about the pendulum's temporal evolution].``If you knew Time as well 
as I do," said the Hatter,``you wouldn't talk about wasting it. It's him." ``I don't know what you mean'' 
said Alice.``Of course you don't !" the Hatter said, tossing his head contemptuously. ``I dare say you 
never even spoke to Time!" ``Perhaps not,' Alice cautiously replied.}

\section{Lagrange}

Lagrange has found a formulation of Mechanics that makes it possible to free oneself from the 
complicated drawings asked to our high school students (Fig. \ref{fig1}) in order to solve a problem in 
Mechanics by identifying the different forces at play, finding their resultant, and then solving Newton's 
equations. Unlike Newton, who tried to present his results in the most ancient form possible, as Euclid 
(around 300 ACN) would have liked, Lagrange understood the importance of no longer having to 
make``inspired drawings" to have a chance of solving a mechanical problem. Moreover, if Newton's 
equations are simple in Cartesian coordinates, they become much more complicated in other coordinate 
systems better suited to the geometry of the system under study, as polar coordinates in the FP case.

By invoking a ``principle of least action", Lagrange replaced Newton's technical arsenal by the search for 
a single function, named after him \textit{Lagrangian} and denoted $L$. Typically, $L=E_k-E_p$, which is 
the difference between the kinetic and potential energies of the system. Lagrange's postulate is that the 
system, during its temporal evolution between two instants $t_1$ and $t_2$, will always minimise the 
action, $S=\int_{t_1}^{t_2} L\, dt\,$. 
The Lagrangian can be written in any coordinate system and the real number 
it provides for a given state of the system will always be the same, 
regardless of the coordinates used. For example the FP Lagrangian is 
$L=\frac{m}{2}\left(\dot r^2+r^2 (\dot \theta+\Omega)^2 \right)-\frac{m}{2}\ \omega^2 r^2\,$, 
where $(\dot r, \dot\theta)$ are the rates of changes of $(r,\theta)$. The space parameterised by 
the Lagrange variables $(r,\theta)$  is called \textit{configuration space}.

From the principle of least action, equations of motion can be deduced which take the same form in all coordinate systems and are, in the present case of the FP, equivalent to Newton's equations: 
$\frac{\partial L}{\partial q_a}-\frac{d}{dt}\frac{\partial L}{\partial \dot q_a}= 0\,$, with $q_a$ the position 
coordinates. The symbols $d$ and $\partial$ denote the total and partial derivatives, respectively. 
But that is not all. 
Lagrange's analytical formalism also makes it possible to ignore the vast majority of forces at play, 
such as the tension in the cable of the FP. It is therefore of little importance to know the tensions and 
constraints imposed on the solid body whose movement is being studied. All that matters is to know 
that the body remains solid and subject to external constraints such as the weight $\vec P\,$ 
and the fact that the length of the FP cable is constant. 

A detailed discussion of the FP in Lagrangian and Hamiltonian formalisms can be found in \cite{saletan}. 
In the rest of this article, we will take inspiration from this book, extracting the key concepts while 
limiting the associated mathematical developments.

\section{From Earth to phase space}

The choice of a ``good" coordinate system reflects the \textit{symmetries} of the system under study. 
Hamilton sought for even more symmetry by not limiting the visualization of a system to position 
variables only. In his formalism he introduced from the very beginning the following fundamental 
idea: the variables of position and velocity are independent. They can therefore be chosen 
independently at any moment and the dynamical equations ``propagate" these initial conditions into the 
future. Hamilton defined variables named \textit{momenta} $P_a=\frac{\partial L}{\partial \dot q_a}$ 
generalising the velocities (one for each position coordinate) and then constructed, for each 
Lagrangian $L\,$, a new function $H = P_a \dot q_a-L\,$ named \textit{Hamiltonian}. 
The space spanned by the independent position and momentum variables is called phase space. 
Using the principle of least action extended to phase space, Hamilton obtained first-order differential 
equations, $\dot P_a=-\frac{\partial H}{\partial q_a}\,$ and $\dot q_a=\frac{\partial H}{\partial P_a}\,$, unlike Newton's and Lagrange's equations which are of the second order in the time-derivatives. 
The number of these first-order equations is equal to the phase space dimension. 
One half of Hamilton's equations allows (in the cases encountered in solid body Mechanics) 
to express the momenta as functions of the velocities while the other half gives equations which, 
combined with the first half, reduce to the Lagrange equations.

The Hamiltonian for the FP is of the form 
$H=\frac{P_r^2}{2m}+\frac{P_\theta^2}{2mr^2}-\Omega \,P_\theta+\frac{m}{2}\, \omega^2 r^2\,$, 
i.e., a function of the position $(r,\theta)$ and momentum $(P_r,P_\theta)$ variables. The space 
$(P_r,r,P_\theta,\theta)$ is the FP phase space. 
Hamilton's equations yield 
\begin{align}
\label{eq3}
P_r=m\dot r\,,\quad   & P_\theta = mr^2 (\dot \theta+\Omega)\,, 
\nonumber \\					 	
\dot P_\theta=0\,,\quad  & \dot P_r=-m\omega^2 r+\frac{P_\theta^2}{mr^3} \;.	
\end{align}
						
They reveal the existence of a constant of motion, $\vert P_\theta\vert=j=m r_0^2 \Omega\,$. 
In the case of Sainte-Waudru's FP, it takes the value  7.14 10$^{-3}$ Js. 
Moreover, Hamilton's equations lead to $\ddot r =-\omega^2 r+\frac{j^2}{mr^3}$ 
and finally to the radial trajectory given by Eq. (\ref{eq1}). 
\begin{center}
	$\backsim$
\end{center}
\textsl{[Alice hadn't seen the Caterpillar enter, too busy meditating on phase space. ``I can't see it," she exclaimed.] ``What do you mean by that?" said the Caterpillar sternly. ``Explain yourself!" ``I can't explain myself, I'm afraid, sir," said Alice, ``because I'm not myself, you see." ``I don't see," said the Caterpillar. ``I'm afraid I can't put it more clearly," Alice replied very politely. [It's quite simple, said the Caterpillar who, exhaling the smoke from his pipe, drew two elegant curves in the air.]
}
\begin{center}
	$\backsim$
\end{center}

\begin{figure}[ht]
	\centering
	\includegraphics[width=0.7\textwidth]{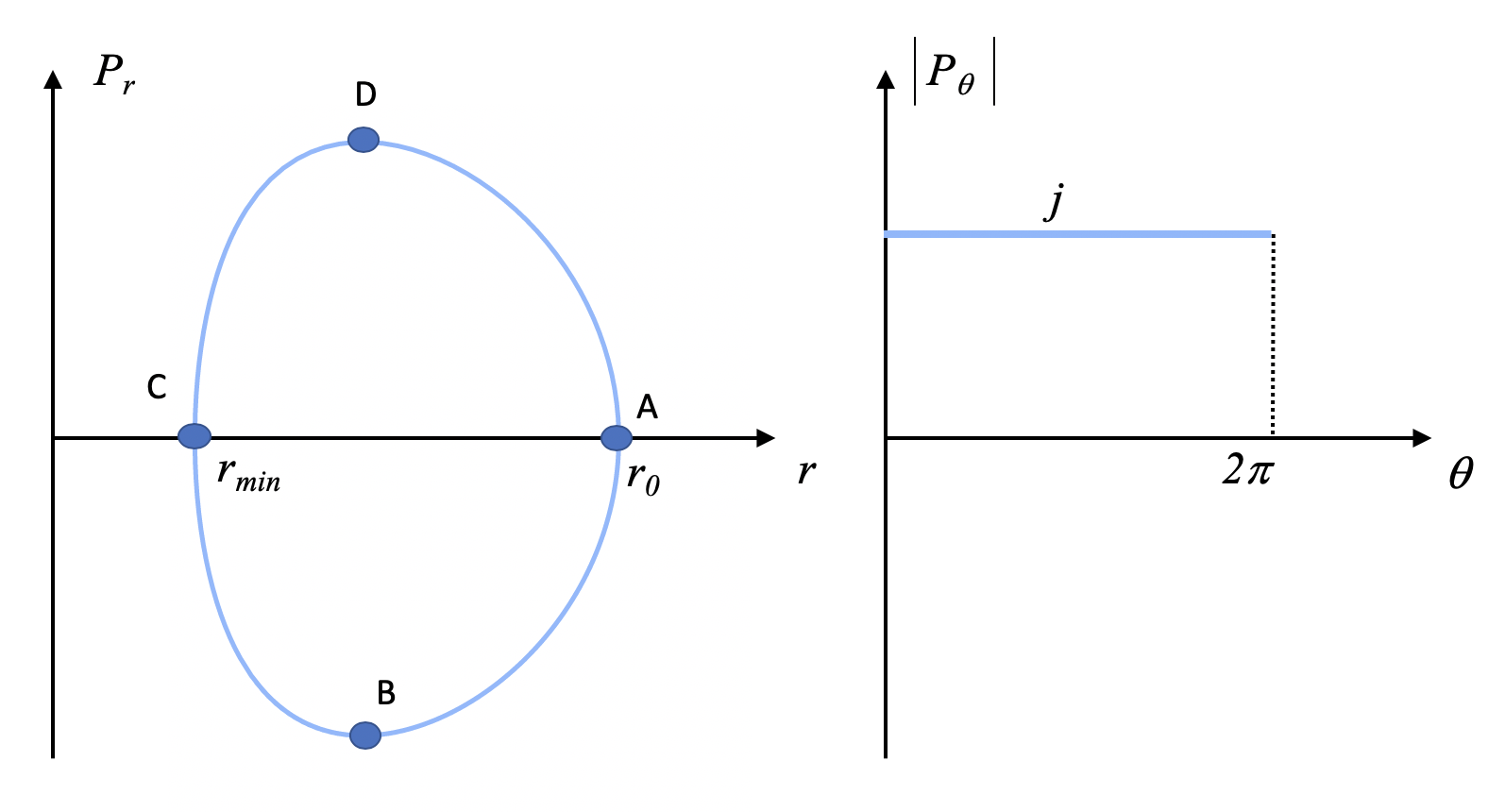}
	\caption{Left: Foucault's pendulum trajectory in the plane $(r ,P_r )$ of phase space, computed with (\ref{eq1}) and (\ref{eq3}). Right: Same for the plane $(\theta ,P_\theta )$.  }
	\label{fig5}
\end{figure}

The dynamics of the FP can be visualised via the pairs of variables $(r,P_r )$ and $(\theta,P_\theta )$, each describing a plane in which the dynamics will draw curves. The phase space provides a succinct geometrical representation of the dynamical evolution of the pendulum, illustrated in Fig. \ref{fig5}. In the plane $(\theta ,P_\theta )$, $P_\theta$ is constant during the complete oscillation. Let us now go through the pendulum's oscillation cycle in the plane $(r ,P_r )$, clockwise from point A. This point 
corresponds to the release of the bob: the amplitude is maximum and the velocity is zero. 
Then the mass falls down, its radial velocity decreases $(P_r<0)$, reaches a minimum in B and 
becomes increasingly close to the local vertical. As previously said, the mass $m$ reaches a minimal 
value of $r$ (in C). Then $r$ increases and $P_r>0$ reaches a maximal value in D. 
The bob of mass $m$ finally returns to point A, from where it starts a new oscillation cycle.

Why introducing this apparent complication created by the doubling of variables between the 
configuration space and the phase space, if one finds Lagrange's equations at the end of the day? 
The reason is that the phase space has a remarkable geometrical structure that allows canonical 
transformations: transformations of the phase-space coordinates that preserve the expression of 
Hamilton's equations of motion. In general, such transformations mix the position and momentum 
variables between them, rendering such a distinction actually artificial. 
The phase space variables can thus mix and give rise to new variables that not only preserve 
the simple expression of Hamilton's equations, but also offer an immediate resolution of these 
equations of motion, as we will now explain. 
\begin{center}
	$\backsim$
\end{center}
\textsl{[Alice, walking and thinking, had arrived at the Duchess's castle.] ``Thinking again?" the Duchess asked, with another dig of her sharp little chin. ``I've a right to think," said Alice sharply, for she was beginning to feel a little worried.}

\section{Strolling on a torus}

\begin{figure}[ht]
	\centering
	\includegraphics[width=0.7\textwidth]{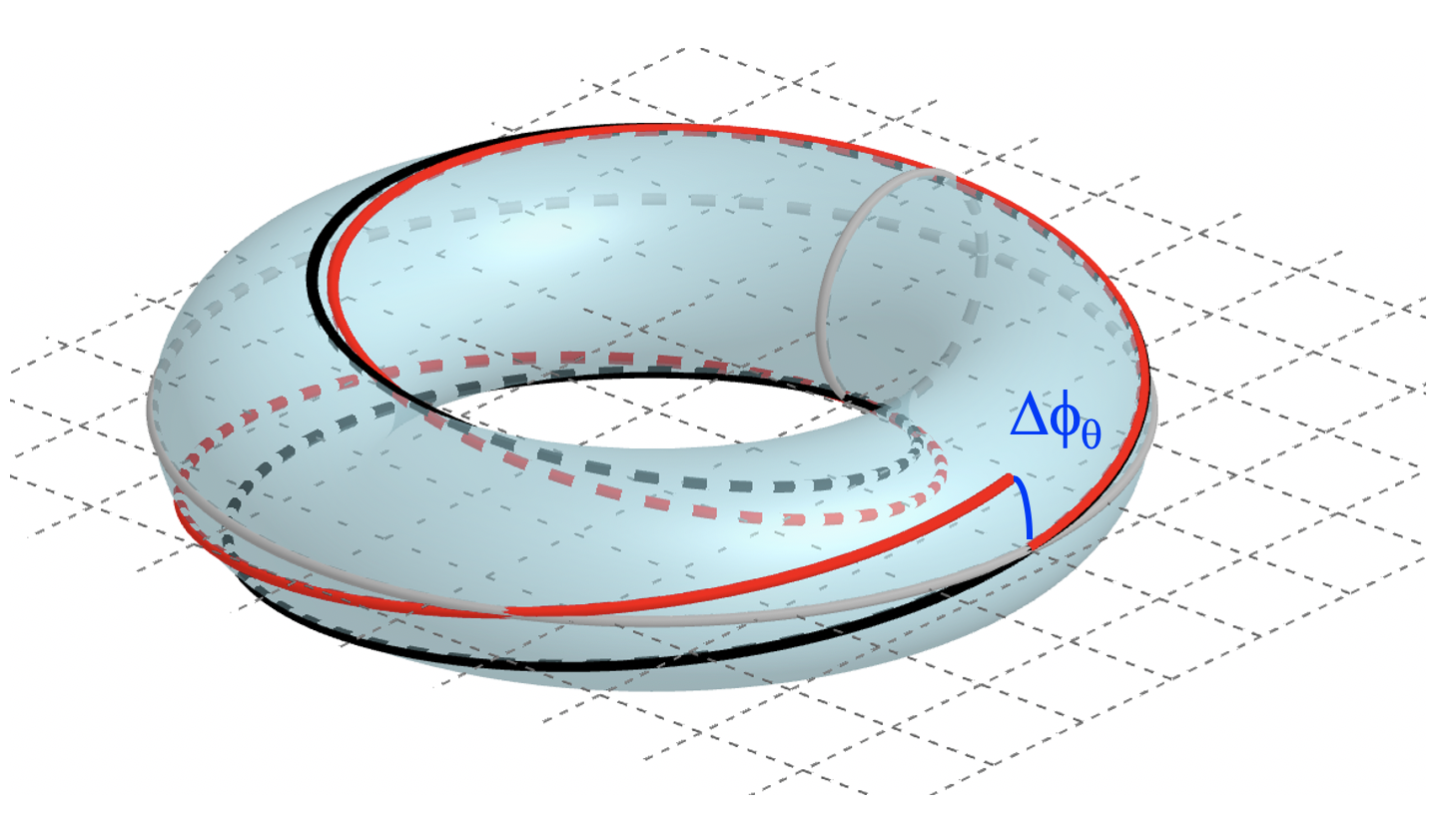}
	\caption{Representation of the phase space of the Foucault pendulum in terms of the angle-action variables.\cite{ggb} 
	The sections of the torus (grey circles) are of dimensions $J_r$ and $J_\theta\,$. 
	The curves on the surface of the torus correspond to the trajectory of the pendulum neglecting 
	the rotation of the Earth (in black) or including it (in red). A point of the curve corresponds to a 
	value of the couple $(\phi_r,\phi_\theta)\,$, i.e., the two angles necessary to be located on the 
	torus. The Hannay angle, $\Delta\phi_\theta\,$, is shown too.  }
	\label{fig6}
\end{figure}

For a bounded motion such as that of the FP, a canonical transformation always exists which makes 
it possible to construct new coordinates starting from $(P_r,r,P_\theta,\theta)$ and leads to 
the action-angle coordinates $(J_r,\phi_r,J_\theta,\phi_\theta)\,$. 
The good surprise is that the action variables $(J_r,J_\theta)$ are invariant over time for given 
initial conditions. 
Some calculations show that $J_\theta=P_\theta$ and \cite{berg}
$J_r=\frac{E+\Omega j}{2\omega}-j/2\,$; 20 Js in our example. 
These are proportional, respectively, to the area of the rectangle shown to the right of Fig. \ref{fig5} 
and the area of the curve shown to the left of that figure. The variables $(\phi_r,\phi_\theta )$ range 
from 0 to $2\pi$ and give the system's position in the cycle of the periodic motion. 
As shown in Fig. \ref{fig6}, the dynamical evolution of the FP in phase space can be seen as the 
displacement of a particle on a torus -- a doughnut. The surfaces of the two sections of the doughnut 
are equal to $J_r\,$ and $j\,$ while the two angles that describe the circulation of the particle on the 
torus are the angle variables.

More generally, for a dynamics which occurs in a compact volume of the phase space, 
a transformation exists that allows to see any dynamical evolution, in the phase space, 
as the displacement of a particle on a torus of dimension $n$, if $n$ is the number of independent 
position variables. 
The surfaces of the different sections of the torus are the action variables while the $n$ angles 
that describe the circulation of the particle on the torus are the angle variables.

Let us come back to FP. The Hamiltonian takes a simple form in action-angle coordinates: 
$H(P_r,r,P_\theta,\theta)\rightarrow H' (J_r,\phi_r,J_\theta,\phi_\theta )=\omega (2J_r+j)+\Omega j\,$. 
Hamilton's equations of motion lead to $\dot \phi_\theta=\frac{\partial H'}{\partial j}=\omega+\Omega$ 
and $\dot \phi_r=\frac{\partial H'}{\partial J_r}=2\omega\,$. 
Would the Earth not rotate, one would have $\Omega=0$ and the angle variable $\phi_r$ would 
increase exactly two times faster than $\phi_\theta\,$. 
In the representation proposed in Fig. \ref{fig6}, 
the complete trajectory of the FP would draw a closed curve (in black). 
However, the Earth rotates and therefore $\Omega$ is different from zero. 
It causes a perturbation in the evolution of $\phi_\theta\,$. The red curve, that corresponds to the actual 
situation, is not closed. There is a discrepancy between the red and black curves and the resulting 
angular deviation is a Berry-Hannay angle for the FP. \cite{Berry1985,Hannay1985}
It is equal to $\Delta\phi_\theta=\Omega T$ and 
corresponds to the angular displacement of the oscillation plane over one period.
The geometric notion underlying the Berry-Hannay angle is that of \emph{holonomy}. 
For extensive discussions on this crucial concept in Physics and for more references, 
see the book \cite{Marsden} and the original papers reprinted in \cite{doi:10.1142/0613}.

\section{Conclusion: The wonderland of Mechanics}

\textsl{``Oh, don't bother me," said the Duchess; ``I never could abide figures." [Alice] had quite 
forgotten the Duchess by this time, and was a little startled when she heard her voice close to 
her ear. ``You're thinking about something, my dear, and that makes you forget to talk. I can't tell 
you just now what the moral of that is, but I shall remember it in a bit." ``Perhaps it hasn't one," 
Alice ventured to remark. ``Tut, tut, child!" said the Duchess. 
``Everything's got a moral, if only you can find it."}
\begin{center}
	$\backsim$
\end{center}

The path that we have made in the different formulations of Mechanics starts from a formulation 
where the fundamental observable is the trajectory of a given solid body as well as its velocity and 
acceleration, obtained by successive time derivations of the position. At the end, the problem is 
represented in terms of the invariants of motion (the action variables) and of the positioning of the 
moving body on the different cycles constituting this motion (the angle variables). The evolution of 
Mechanics thus tends to abstract from the particularities of each problem to evolve towards a general 
formalism, where all dynamical systems with (quasi-)periodic motion can be characterised by 
universal concepts. The Berry-Hannay angle is such a concept: it gives a measure of the 
rotation of the oscillation plane through the non-closed trajectory of the bob of Foucault's pendulum 
in the space of the action-angle variables.

The particular behaviour of the Foucault pendulum can also be interpreted in terms of the parallel 
transport of the velocity vector. This notion has become central in modern Physics through the 
concept called \emph{connection}; here, the Levi-Civita connection on the sphere, 
which expresses how the tangent vectors to the sphere are transported parallel to themselves 
along a given curve on the sphere. This same Levi-Civita concept of parallel transport 
allowed Einstein $(1879-1955)$, with the help of Marcel Grossmann $(1878-1936)$, to better 
understand Gravitation. Indeed, the mathematics of General Relativity is entirely based 
on Levi-Civita's notion of parallel transport on a curved manifold representing our space-time 
populated with matter and radiation.

Finally, the Hamiltonian formalism of Mechanics makes it possible to completely solve many problems 
that cannot be solved otherwise, such as the problem of a massive particle attracted by two centres 
(for example the Earth subject to the combined gravitational attractions of the Sun and Jupiter), 
see e.g. \cite{saletan}.  
Hamilton's approach is instrumental for controlling the position of a satellite or space probe, or for 
understanding the general dynamics of chaotic systems. \cite{Arnold} 
For more references on the wide variety of applications of the Hamilton's formulation of 
classical Mechanics, see \textit{e.g.} the Introduction of the very comprehensive book \cite{Marsden}. 
Finally, Hamilton's formulation is also central in Quantum Mechanics and Quantum Field Theory, 
see e.g. \cite{Weinberg,Srednicki} and references therein.
Nature is quantum at its most fundamental level, hence the crucial importance of the Hamiltonian 
method to describe interacting quantum particles moving at speeds close to the speed of light, 
as is the case with the cosmic rays that traverse us at every single instant.

\begin{center}
	$\backsim$
\end{center}
\textsl{``Wake up, Alice dear!" said her sister; ``Why, what a long sleep you?ve had!" ``Oh, I've had such a curious dream!" said Alice, and she told her sister, as well as she could remember them, all these strange adventures of hers that you have just been reading about (\dots)  So Alice got up and ran off, thinking while she ran, as well she might, what a wonderful dream it had been [and what a wonderful science Mechanics is.]}

\begin{acknowledgments}

We thank Francesco Lo Bue for the successful installation of Foucault's pendulum at Saint Waudru's collegiate church and for fruitful discussions at early stages of this manuscript. We thank Tracy Gittins for his advise on English. 

\end{acknowledgments}

\end{document}